\documentclass[aps,
twocolumn,
]{revtex4} 
\usepackage{epsf}
\usepackage{amssymb,amsmath}
\usepackage{hyperref}
\usepackage{graphicx}

\begin{document}
%
\title{Comparison of the Adsorption Transition for Grafted and Nongrafted Polymers}
\author{Monika M\"oddel}
\email[E-mail: ]{Monika.Moeddel@itp.uni-leipzig.de}
\author{Wolfhard Janke}
\email[E-mail: ]{Wolfhard.Janke@itp.uni-leipzig.de}
\homepage[\\ Homepage: ]{http://www.physik.uni-leipzig.de/CQT.html}
\affiliation{Institut f\"ur Theoretische Physik and Centre for Theoretical Sciences (NTZ),
Universit\"at Leipzig, Postfach 100\,920, D-04009 Leipzig, Germany}
\author{Michael Bachmann}
\email[E-mail: ]{bachmann@smsyslab.org}
\homepage[\\ Homepage: ]{http://www.smsyslab.org}
\affiliation{Center for Simulational Physics, 
 The University of Georgia, \\Athens, Georgia 30602, United States}%

\begin{abstract}
    We compare the thermodynamic behavior of a finite single nongrafted polymer near an attractive substrate with that of a
    polymer grafted to that substrate. 
    After we recently found first-order-like signatures in the microcanonical entropy at the adsorption transition in the nongrafted case,
    and given the fact that many studies on polymer adsorption in the past have been performed for grafted polymers,
    the question arises, to what extent and in what way does grafting change the nature of the adsorption transition? 
    This question is tackled here using a coarse-grained off-lattice polymer model and covers not only the adsorption transition but also all other transitions
    a single polymer near an attractive substrate of varying strengths undergoes.
    Because of the impact of grafting especially on the translational but also on the conformational entropy of desorbed chains, the adsorption transition
    is affected the strongest. Our results are obtained by a combined canonical and microcanonical analysis of parallel tempering Monte Carlo data.
\end{abstract}
\maketitle

\section{Introduction}
    The conformational properties of a polymer are characterized by a collective, cooperative be\-havior of the monomers in response to different system conditions.
    The geometric constraint of an attractive substrate adds to the freezing and coil-globule transitions of a polymer in bulk solution 
    the adsorption transition and gives rise to several phases induced by the competition between monomer-monomer and monomer-surface attraction.
    This behavior of grafted polymers on substrates is of manifold practical importance. 
    Ultrathin end-grafted polymer layers play a major role in adhesion, colloidal stabilization\cite{russel,chevigny,gast},
    chromatography\cite{chromatography}, lubrication, microelectronics, and biocompatibility of artificial organs. 
    To achieve the grafting, typically ``grafting from'', ``grafting through'' or ``grafting onto'' polymerization techniques are used\cite{graftingonto,grafting1,grafting2} 
    or diblock copolymers are physisorbed to the substrate\cite{diblock}.

    Grafted polymer adsorption is computationally easier to handle since the phase space lacks potentially 
    desorbed conformations some distance away from the substrate. Additionally, one avoids the introduction of a
    hard wall parallel to the surface that is necessary to prevent the polymer from escaping but introduces a further parameter
    that might not always be of interest. 
    This is probably the main reason for the prevalence of grafted polymers in the theoretical studies on polymer 
    adsorption\cite{huang,descas,binder1,Luettmer-Strathmann,Metzger,prellberg,kumar}. 
    Also the adsorption of nongrafted polymers has been studied\cite{mbhomo1,mbhomo2,wang},
    but usually those works have been performed on different models which hamper the extraction of the influence of grafting on the results by comparison.

    It is the goal of our study to fill this apparent gap and systematically analyze and compare the phase behavior of a single grafted homopolymer chain model
    with that of a nongrafted but otherwise identical model.
    The model assumes an implicit solvent and an attractive substrate and in the case of the nongrafted chain a sterical wall some distance away from the attractive substrate
    confining the polymer. Hence, also the nongrafted polymer is not completely free. Nevertheless, we use the terms ``nongrafted'' and ``free''
    synonymously in the following.
    Our comparison spans a wide range of temperatures and surface attraction strengths.

\section{Model and Methods}
    Our study focuses on a simple ``bead-stick'' model of a linear polymer with fixed bond length
    and with three terms that contribute to the energy\cite{Moeddel2009,Moeddel2010},
        \begin{eqnarray}\label{eq:energy}
        E & = & 4\sum_{i=1}^{N-2}\sum_{j=i+2}^{N}\left( {r_{ij}^{-12}}-{r_{ij}^{-6}}\right)+
        \frac{1}{4}\sum_{i=1}^{N-2}\left( 1-\cos\vartheta_{i}\right)\nonumber\\
        & & +\, \epsilon_{s}\sum_{i=1}^{N}\left(\frac{2}{15} {z_i^{-9}} - {z_i^{-3}}\right),
        \end{eqnarray}
    where the first two terms are the energy of a polymer in bulk that consists of the standard 12-6 Lennard-Jones (LJ) potential
    and a weak bending energy. The distance between the monomers $i$ and $j$ is $r_{ij}$, and $0\leq \vartheta_i \leq \pi$ denotes the
    bending angle between the $i$th, $(i+1)$th, and $(i+2)$th monomer. The third term is the attractive surface potential,
    where $z_i$ is the distance of the $i$th monomer to the substrate. 
    It is obtained by integration over the continuous half-space $z<0$, where every space element interacts with a single monomer by the 
    12-6 LJ expression\cite{steele}. 
    The adsorption strength is controlled by 
    the parameter $\epsilon_s$ which weighs the monomer-surface and monomer-monomer interaction. 

    This model is chosen to facilitate comparison with previous work done by our group\cite{Moeddel2009,Moeddel2010,christoph,christoph2} and adapted from the AB model\cite{stillinger1,stillinger2},
    a standard model for hydrophobic-polar representations of
    peptides. In fact, the bending stiffness is a remnant of this modelling that has no qualitative influence on the results of this work.

    Besides the temperature $T$, we consider the adsorption strength $\epsilon_s$ as a control parameter and construct the  
    phase diagram in the $T$-$\epsilon_s$ plane which reflects the structural behavior of classes of polymers.
    We consider the two cases, where (a) the polymer is grafted with one end to the substrate and (b) where it is allowed to move freely in the space between the substrate and 
    a hard wall a distance $L_z=60$ away from the substrate. The chain length is chosen to be $N=40$.
    The motivation of this comparison is strongly rooted in our findings of Ref.\ \cite{Moeddel2010}, where we observed for a nongrafted 20-mer a strong, but trivial,
    dependence of the translational entropy of desorbed chains on $L_z$. This led us to expect a qualitative difference between grafted and nongrafted adsorption that we 
    subsequently analyze in the present work. 
    As discussed in more detail below,
    it turns out that not only the translational but also the conformational entropy are responsible for the differences.

    Although most of the transitions of such a polymer are related to thermodynamic phase transitions, for finite lengths
    they are typically just smoothly signaled in the fluctuations of canonical expectation values.
    Additionally, those peaks differ in position for different observables,
    such that no well-defined transition point exists for finite systems. 
    Nevertheless, all transitions can be identified and found in a canonical as well as in a 
    microcanonical analysis such that our method comprises both. 
    Dependent on the transition, one or the other ensemble provides a clearer signal of the phase transition. 
    Continuous transitions are sometimes hardly visible in the microcanonical entropy, but they are pronounced in canonical expectation values like
    the specific heat. Additionally, the canonical ensemble seems to be more intuitive since most experiments are performed at constant temperature in a
    heat bath. On the other hand, the microcanonical approach provides a clear and easy method to differentiate first-order-like phase transitions
    from continuous ones and is in a way more fundamental, since it is based on the temperature-independent density of states $g(E)$.
    Hence, a complementary analysis of both approaches provides a more detailed picture of the phase behavior.

    In the canonical ensemble the system is assumed to be in equilibrium with a heat bath 
    of temperature $T$ that it can exchange energy with. 
    All possible states of energy $E$ are distributed according to the Boltzmann probability $g(E) e^{-\beta E}$, $\beta=1/k_B T$ ($k_B\equiv 1$ in the following),
    and to describe phase transitions, temperature fluctuations of canonical expectation values 
    $\left<O\right>_\beta=\int O(E)g(E)e^{-\beta E} dE/\int g(E)e^{-\beta E} dE$ are investigated. 
    All information encoded in those fluctuations is of course also present in the microcanonical quantities $O(E)$ and $g(E)$. 
    Since fluctuations are integrated out, some information may be smeared out in the canonical analysis, and thus 
    it is worthwhile to also look at these microcanonical quantities 
    as it was, e.g., already successfully done for peptide aggregation\cite{christoph,christoph2}. 
    The microcanonical entropy $S(E)=\ln g(E)$ includes all entropic and energetic information. 
    Phase transitions can be identified via its slope $\beta(E)=T^{-1}(E)=[\partial S(E)/\partial E]_{N,V}$, the inverse microcanonical
    temperature, and its curvature $[\partial\beta(E)/\partial E]_{N,V}$~\cite{gross1,jankeequalheight,gross2,Stamerjohanns}. In the former 
    they are typically signaled by inflection points that yield maxima in the latter\cite{gamma}. 
    Any conformational quantity $O(E)$, e.g., tensor components of the radius of gyration, the distance of the center-of-mass of the polymer
    to the substrate, or the number of attached monomers, adds structural information to the picture. 

    A word of caution is in order here concerning the term ``phase transition''. Phase transitions in the strict thermodynamic sense are only defined for infinite system size.
    Only there the peak positions in the temperature derivatives of different canonical observables
    fall onto the same point, and the microcanonical and canonical ensembles are equivalent for sufficiently short-ranged interactions. 
    Hence, the ``phase transitions'' of the finite system
    described here are not phase transitions in the strict thermodynamic sense and are not uniquely located in the phase diagram.
    They indeed may even differ in nature from the corresponding
    infinite-system phase transition. The term ``phase transition in a finite system'' has to be read in the following with this in mind.

    \begin{figure*}[t!]
        \begin{center}	
            \includegraphics[width=11.9cm]{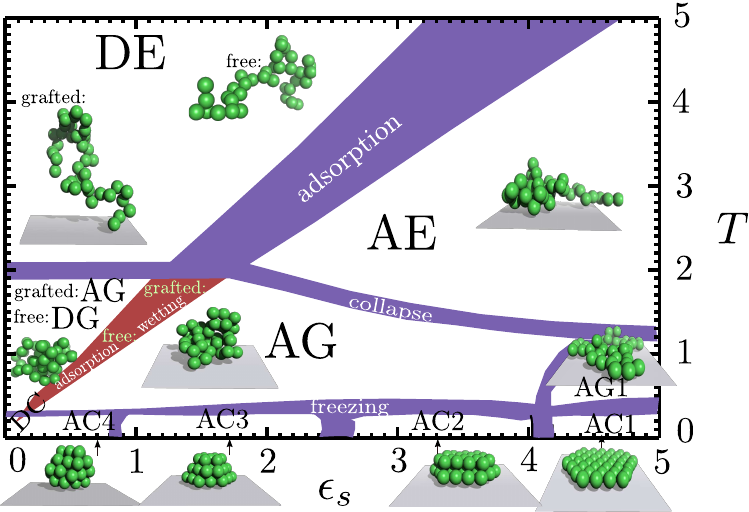}
        \end{center}
            \caption{\label{fig:diagram}The pseudophase diagram in the canonical plane, parametrized by temperature $T$ and adsorption strength $\epsilon_s$, 
            for the 40-mer. The purple transition regions have a broadness
            that reflects the variation of the corresponding peaks of the temperature derivatives of different canonical expectation values we investigated.
            Phases with an ``A/D'' are adsorbed/desorbed. ``E'', ``G'' and ``C'' denote phases with increasing order: expanded, 
            globular, and compact/crystalline, where the compact phase occurs with different numbers of layers. The AG phase is divided into a 
            phase of planar globular conformations for high surface attraction strength (AG1) 
            and one with a significantly higher extension perpendicular to the substrate (AG). The main difference between
            grafted and free chains occurs at the adsorption transition.} 
    \end{figure*}

    To study the model (\ref{eq:energy}), the parallel-tempering method\cite{paralleltemp1,paralleltemp2} together with 
    an error-weighted histogram reweighting method similar to the one discussed in Ref.~\cite{fenwick} was used.
    64-72 different replicas 
    covering uniformly a temperature range from $T=0.001$ to $T=50$ 
    were chosen with 50\,000\,000 sweeps each, from which every 10th value was stored in a time series -- the autocorrelation time 
    is of the order of thousands of sweeps. For the canonical analysis this statistics is very generous (about a fifth would have sufficed to gain a decent 
    overview of the behavior),
    but to also reduce the fluctuations of the observables versus energy, in particular around the freezing transition, we increased the statistics. 
    For low surface attraction strength, 64 replicas were enough to guarantee a sufficient overlap of the energy histograms. At higher $\epsilon_s$, the 
    low-temperature histograms have less overlap such that the addition of a very few extra replicas became necessary.
    All energy histograms $H_i(E)$ from the parallel tempering run at inverse temperatures $\beta_i=1/T_i$ can be reweighted to yield an estimate of the density
    of states $g_i(E)\propto H_i(E)e^{\beta_i E}$. These estimates are only of reasonable quality in an energy regime with sufficient statistics,
    i.e., where the canonical energy distribution is peaked.
    Since absolute estimates of the partition function cannot be obtained in importance sampling Monte Carlo simulations, 
    there is an unknown prefactor that is different for every $\beta_i$. To get rid of it, we work with the ratio $g_i(E+\Delta E)/g_i(E)$ 
    and since the density of states spans many orders of magnitude, it is advantageous to use the logarithm 
        \begin{eqnarray}
            \Delta S_i(E) &=&S_i(E+\Delta E)-S_i(E)\\
            &=&\ln\left[g_i(E+\Delta E)/g_i(E) \right] \nonumber\\
            &=&\ln\left[H_i(E+\Delta E)\right]-\ln\left[H_i(E)\right]+\beta_i\Delta E.\nonumber
        \end{eqnarray}
    Now, a weighted average over all histograms 
        \begin{eqnarray}
            \overline{\Delta S}(E)&=&\frac{\sum_i\Delta S_i(E) w_i(E)}{\sum_i w_i(E)}
        \end{eqnarray}
    can be taken with a weight approximately inversely proportional to the variance 
        \begin{eqnarray}
            w_i(E)=\frac{H_i(E+\Delta E)\cdot H_i(E)}{H_i(E+\Delta E)+H_i(E)}\propto 1/\sigma^2(\Delta S_i(E)).
        \end{eqnarray}\\
    This gives an excellent estimate of $\beta(E)\approx \overline{\Delta S}(E)/\Delta E$ as long as all histograms 
    overlap -- which parallel tempering requires anyway. Up to a constant, the microcanonical entropy $S(E)$ is obtained by integration.
    Conveniently, apart from being easily implemented and of similar performance as the more established multiple histogram
    reweighting or weighted histogram analysis methods (WHAM)\cite{ferrenbergswendsen,wham}, 
    this reweighting method directly estimates $\beta(E)$, which is the starting point of our
    microcanonical inflection point analysis\cite{gamma}.

    Estimates of the statistical uncertainties are obtained by means of the jack-knife method\cite{efron}, which yields robust and almost unbiased results even when
    applied to non-linear analyses of the raw data (which, in our case, are the time series for each of the parallel-tempering replica). 
    Our statistics is so high that the error bars are hardly visible on the scale of the following figures. For better readability we have, therefore, included them
    only in a few representative cases. 

    In the following, we will collect and compare the findings for the different transitions.

\section{Results}\label{sec:results}
    All transitions are contained in the phase diagram in Figure \ref{fig:diagram}. 
    It is constructed by combining the information encoded in the canonical expectation values and their temperature derivatives for a number of observables. 
    To give an example, 
    the freezing transition is signaled by a pronounced peak in the specific heat, $c_V(T)$, the temperature derivative of the canonical expectation value of the energy
    given in eq.\ (\ref{eq:energy}). A peak in the temperature derivative of the canonical expectation value of the squared radius of gyration, $R_{\rm gyr}^2= \sum_{i=1}^N\left(\vec{r}_i-\vec{r}_{\rm cm}\right)^2/N$, with 
        $\vec{r}_{\rm cm}=(x_{\rm cm},y_{\rm cm},z_{\rm cm})=\sum_{i=1}^N \vec{r}_i/N$ being the center-of-mass of the polymer,
    indicates the collapse transition quite clearly. Also, its tensor components parallel and perpendicular to the substrate, 
        $R_\parallel^2=\sum_{i=1}^N[\left(x_i-x_{\rm cm}\right)^2+\left(y_i-y_{\rm cm}\right)^2]/N$ and 
        $R_\perp^2=\sum_{i=1}^N\left(z_i-z_{\rm cm}\right)^2/N$, offer rich information.
    The adsorption transition can best be seen by the number of surface contacts, where a monomer is defined to 
    be in contact with the surface
    if $z_i<1.5$, or by the distance of the
    center-of-mass of the polymer to the substrate. 
    Some observables exhibit a peak at all of the transitions, while others are only sensitive to a few of them. 
    For our finite system, at each transition, the peaks of different observables occur at slightly different temperatures and have a certain width.
    The widths of the transition ``bands'' in Figure \ref{fig:diagram} approximately cover the regime 
    of those different transition peaks.

    The phase diagram in Figure \ref{fig:diagram} generalizes the one presented in 
    Ref.~\cite{Moeddel2009}, since the grafted case is considered as well. 
    A strong difference is observed at the adsorption transition, while the other transitions remain more or less unaffected. 
    The key aspect is by how much the grafting affects the two phases involved in any of the transitions at hand.\\

    \begin{figure}[t!]
        \begin{center}	
            \includegraphics[width=8.2cm]{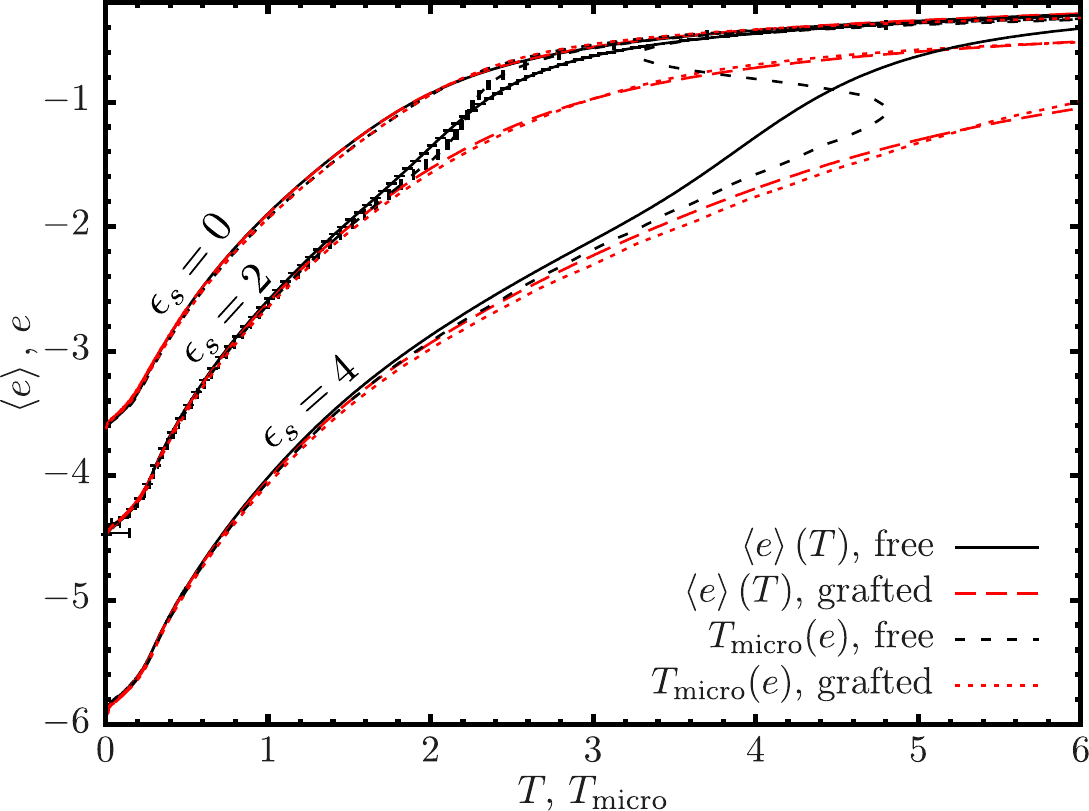}
        \end{center}
            \caption{\label{fig:evsT}Canonical expectation values of the energy $\left<e\right>(T)$ versus temperature $T$ for three exemplified values of $\epsilon_s$.
                    For low temperatures, grafted and nongrafted chains are adsorbed for $\epsilon_s=2,4$, such that $\left<e\right>(T)$ is very similar. For high temperatures, 
                    $\left<e\right>(T)$ converges to the common average energy of conformations away from the influence of the substrate for nongrafted chains while remaining lower for
                    grafted polymers. For convenience also the corresponding microcanonical quantities are shown (same data as in Figure \ref{fig:together_s_t}(b)).} 
    \end{figure}
\begin{figure}[t!]
        \begin{center}	
            \includegraphics[width=8.3cm]{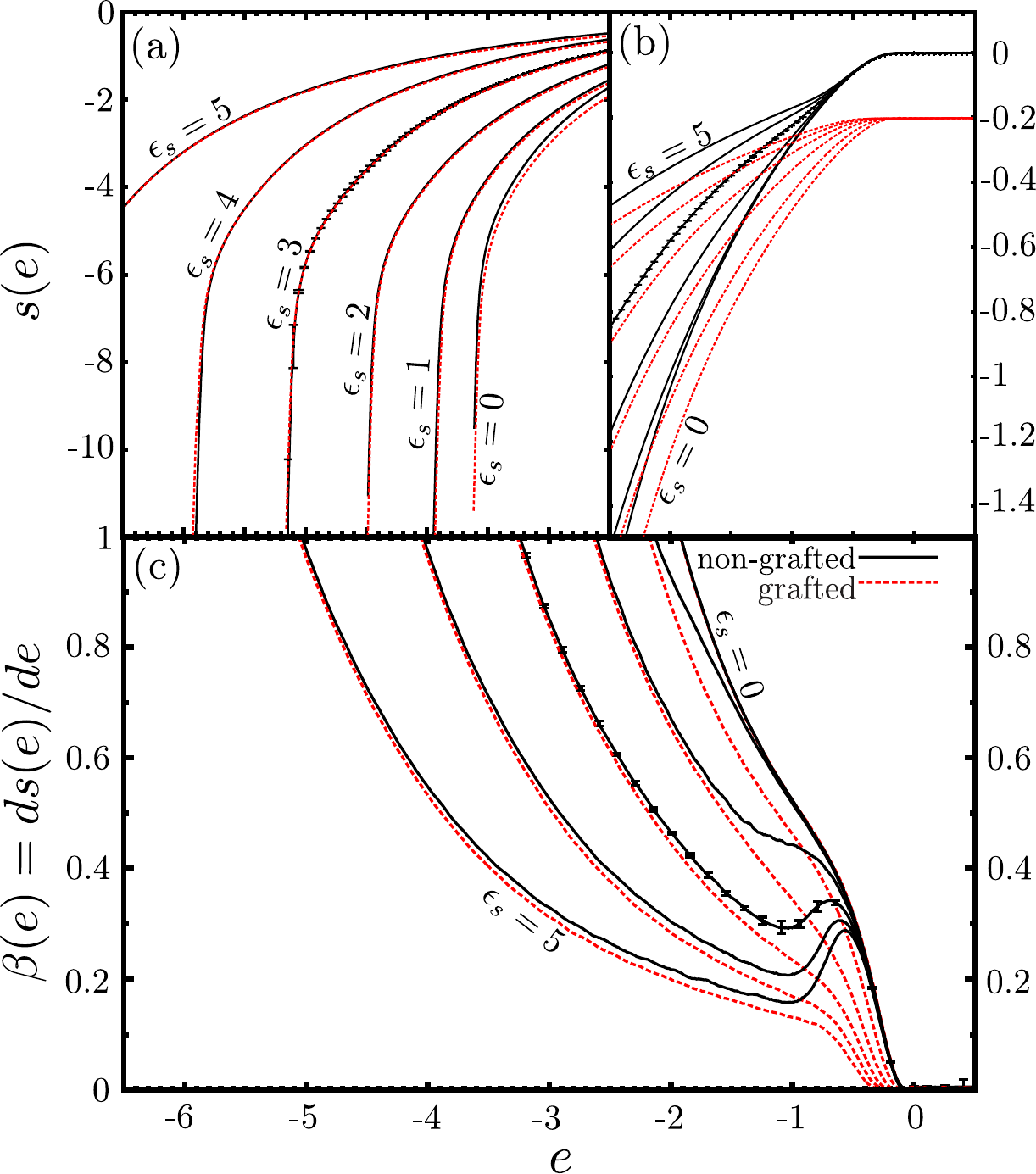}
        \end{center}
            \caption{\label{fig:together_s_t} (a) and (b) both depict the microcanonical entropy $s(e)= \ln g(e)/N$ in different resolutions and span together the energy regime $e\in[-6.5,0.5]$.
            (c) shows its derivative $\beta(e)=\left[\partial s(e)/\partial e\right]_{N,V}$ over the same regime although the $\beta$-values at low energies are not visible.
            For $\epsilon_s=3$, jackknife errors are included for the nongrafted case that are of the same order of magnitude for all other data.
            }
    \end{figure}

\textbf{Energy versus Temperature. }
    Let us discuss first how the canonical expectation value of the energy per monomer $\left<e\right>(T)$ is influenced by the grafting.
    In Figure \ref{fig:evsT},
    additional to the canonical expectation value $\left<e\right>(T)$, also the microcanonical temperature $T_{\rm micro}(e)=[\partial s(e)/de]_{N,V}^{-1}$
    is plotted with exchanged axes. This illustrates the nonequivalence of the canonical and microcanonical ensemble for finite systems. The microcanonical data
    are the same as depicted in Figure \ref{fig:together_s_t}(b) and will be explained in more detail there. For $\epsilon_s=4$, $T_{\rm micro}(e)$ is not bijective 
    anymore such that an additional analysis of the microcanonical data gets interesting here.
    %
    Certainly, $\left<e\right>(T)$ decreases with $\epsilon_s$, since 
    the attractive surface potential becomes more negative with increasing $\epsilon_s$.
    For $\epsilon_s=0$, the grafting has hardly an effect on 
    the energy such that the corresponding curves in Figure \ref{fig:evsT} almost coincide. For larger values of $\epsilon_s$, a crossover occurs from low temperatures,
    where the polymer is adsorbed and the energy of free and grafted polymers is very similar, to high temperatures, where the energy of free polymers approaches the 
    $\epsilon_s$-independent values of polymers in bulk solution while that of grafted polymers is always reduced due to the proximity to the attractive substrate.

    Also visible is the freezing transition as an inflection point close to $T\approx 0.3$ that does not differ much for grafted and nongrafted chains. 
    The collapse cannot easily be identified in the energy plots.
    Figure \ref{fig:evsT} is also supposed to serve as an orientation to compare microcanonical observables plotted versus normalized energy $e=E/N$ with canonical expectation 
    values plotted versus temperature $T$.\\

\textbf{Effect on the Translational Entropy. }
    While for $\epsilon_s=0$, the energy is hardly influenced by the grafting, the same does not hold true for the entropy. 
    Figure \ref{fig:together_s_t}(a) shows that the microcanonical entropy $s(e)$ is not noticeably influenced by the grafting at low energies where in many cases
    the chains are adsorbed. One should keep in mind here that $s(e)$ is only known up to a constant. This constant is chosen in such a way in 
    Figure \ref{fig:together_s_t}(a) and (b) as to overlap $s(e)$ for fixed $\epsilon_s$ at low $e$. 
    The number of states of the grafted polymer should in most cases be smaller than for the free one. Close to the ground state, the difference, however, gets small
    such that this choice is reasonable. In Figure \ref{fig:together_s_t}(b), $s(e)$ for grafted and nongrafted chains separates until for large energies they are separated by a fixed distance given by the additional translational entropy of the free chain proportional to the logarithm of the box size.
    This introduces a box size dependence of the adsorption and was already discussed for the free chain in Ref.\ \cite{Moeddel2010}. 
    The energy regime over which the $s(e)$ curves separate is the energy regime where the polymer desorbs such that this transition is most severely
    affected by the grafting. 
    For $\epsilon_s>2$ the increase in $s(e)$ during the desorption of free polymers even gets large enough to induce a convex regime in $s(e)$, even
    though also the conformational entropy plays its role. Here, at least two phases coexist.
    This is more clearly reflected in the backbending region of $\beta(e)=T^{-1}_{\rm micro}(e)$ as shown in Figure \ref{fig:together_s_t}(c). 
    In the coexistence regime an increase in energy leads to a decrease in temperature.
    This thermodynamically unstable behavior is also signaled by the negativity of the microcanonical specific heat 
    $c_V(e)=[\partial T(e)/\partial e]^{-1}_{N,V}=-[(\partial s/\partial e)^2/(\partial^2 s/\partial e^2)]_{N,V}$. 

    If the system size approaches infinity, the convex intruder in $s(e)$ vanishes such that its derivative $\beta(e)$ is constant over a range of energy that 
    corresponds to the latent heat for first-order phase transitions. For continuous transitions the range of the intruder and the latent heat go to zero. 
    In accordance with theoretical predictions\cite{binder1}, the latent heat at the adsorption transition was found to disappear for long chains in Ref.~\cite{Moeddel2010}.
    With only a very restricted translational entropy, no such convex intruder in $s(e)$ was found in any grafted polymer adsorption.\\

    \begin{figure}[t!]
        \begin{center}	
            \includegraphics[width=8.3cm]{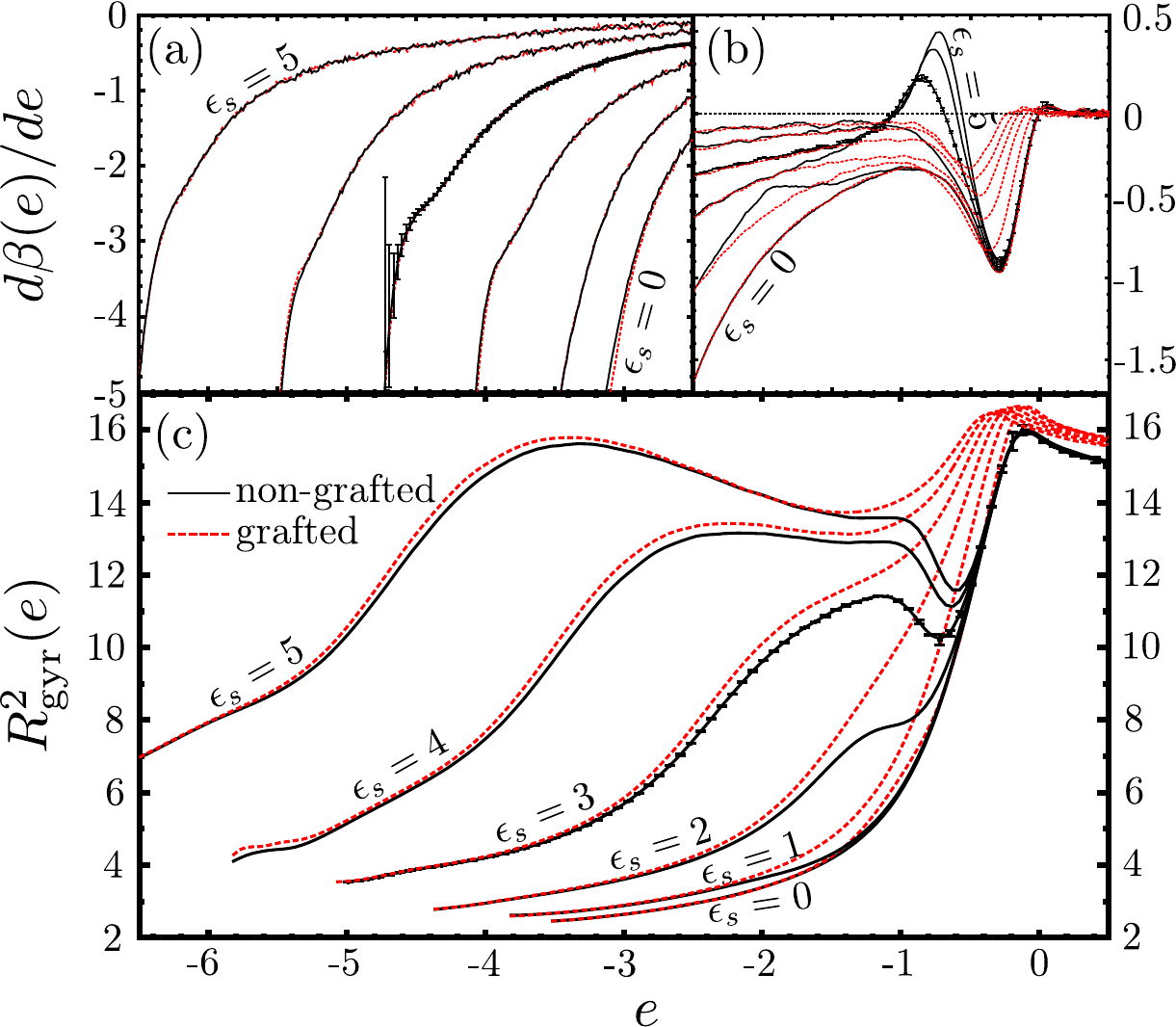}
        \end{center}
            \caption{\label{fig:together_beta_r} (a) and (b) go one 
            derivative further than Figure \ref{fig:together_s_t}(c) and show $\left[\partial \beta(e)/\partial e\right]_{N,V}=\left[\partial^2s(e)/\partial e^2\right]_{N,V}$ again in different resolutions. 
            (c) provides additional structural information via the radius of gyration. The error bars shown for the nongrafted case with $\epsilon_s=3$ are 
            representative for all other data.
            }
    \end{figure}

\textbf{Effect on the Conformational Entropy. }
    The grafting not only forces the polymer to stay close to the substrate and significantly reduces the translational entropy that way but also has an influence
    on the conformation. How strong this influence is can quite nicely be seen in Figure \ref{fig:together_beta_r}(c), where the squared radius of gyration 
    $R_{\rm gyr}^2$ 
    is plotted versus energy. 
    $R_{\rm gyr}^2(e)$ is obtained by only summing over conformations with energies close to $e$.
    Below the collapse transition the radius of gyration -- a measure for the average extension of the polymer -- is almost unaffected. 
    Independent of whether or not the polymer is grafted, a compact shape is attained here with a deformation determined by the strength of the surface potential.

\begin{figure}[t!]
        \begin{center}	
            \includegraphics[width=8.4cm]{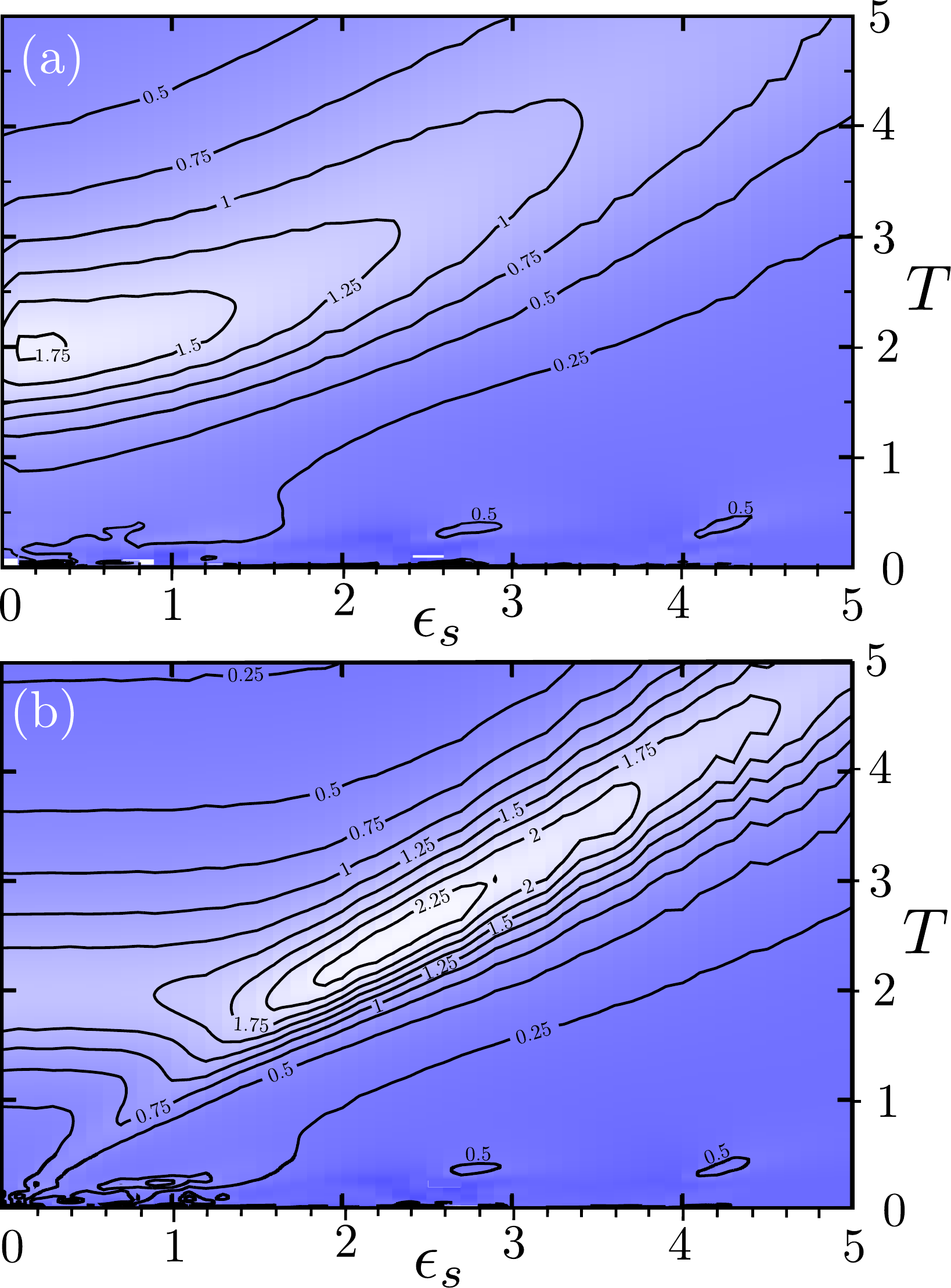}
        \end{center}
            \caption{\label{fig:rgyrz}Fluctuation of the tensor component of the radius of gyration perpendicular to the substrate $d\left<R_{\rm gyr, \perp}^2\right>/dT$ 
                    for (a) the grafted and (b) the free polymer as a contour plot versus surface attraction strength $\epsilon_s$ and temperature $T$.} 
    \end{figure}

\begin{figure}[t!]
        \begin{center}	
            \includegraphics[width=8.15cm]{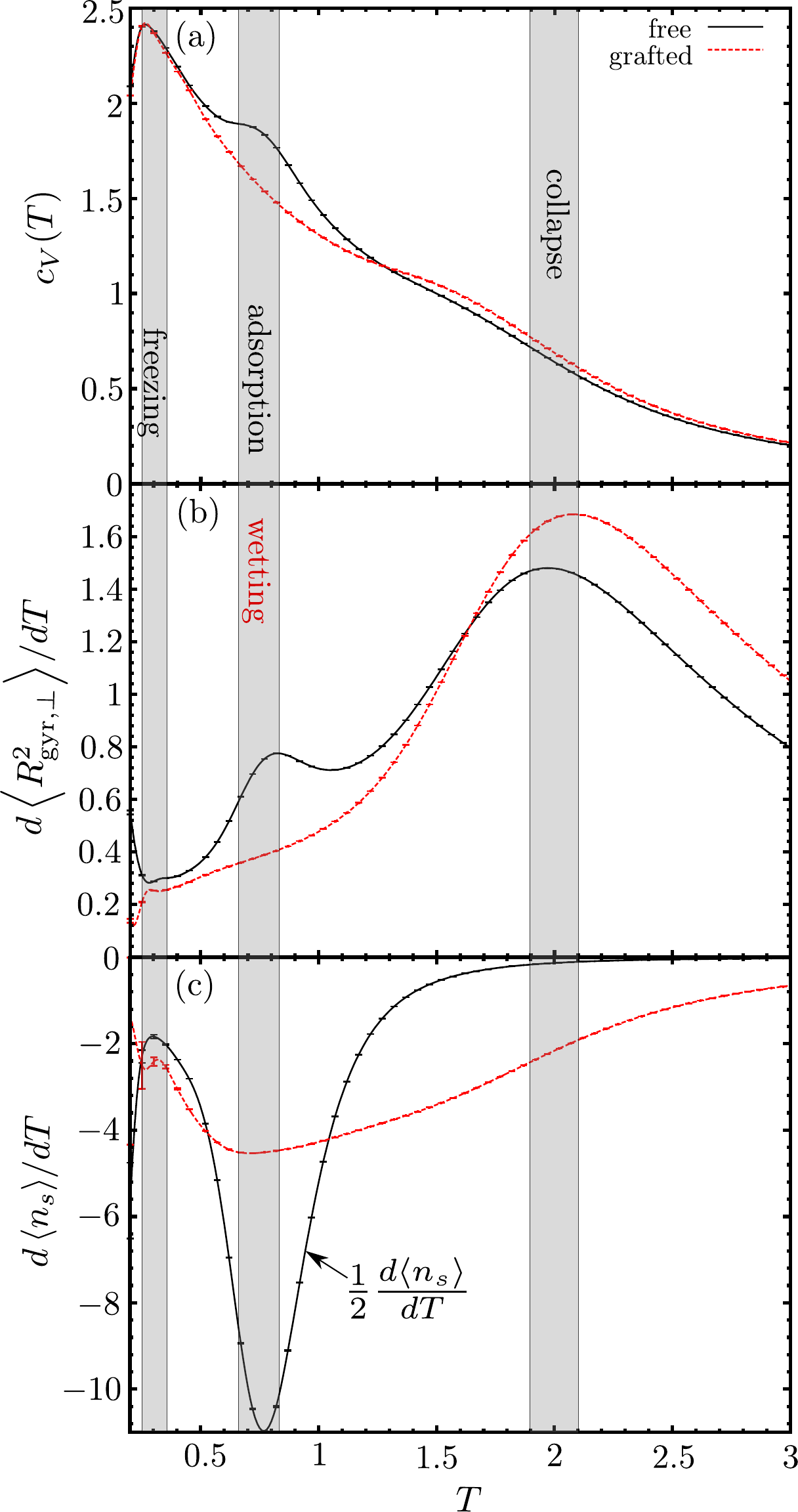}
        \end{center}
            \caption{\label{fig:es0.7} Fluctuations of canonical expectation values for weak surface attraction, $\epsilon_s=0.7$, where the adsorption occurs at a lower temperature
                than the collapse: (a) specific heat $c_V(T)$, (b) the fluctuation of the radius of gyration component perpendicular
                to the substrate $d\left<R_{\rm gyr,\perp}\right>(T)/dT$, and (c) the fluctuation of the number of monomers in contact with the substrate 
                $d\left<n_s\right>(T)/dT$.
                While for the free polymer the adsorption is signaled in all three observables, for a
                grafted one only an activity in (c) is visible, indicating a change from the adsorption into the wetting transition.} 
    \end{figure}

    The \textit{collapse transition} is rather weakly signaled in the microcanonical quantities. 
    For $\epsilon_s=0$, it can be identified
    with an inflection point in $\beta(e)$ in Figure \ref{fig:together_s_t}(c) that directly corresponds to a maximum in $d\beta(e)/de$ in Figure \ref{fig:together_beta_r}(b) close to 
    $e=-1$. As it has to be, at the same energy the squared radius of gyration $R_{\rm gyr}^2(e)$ in Figure \ref{fig:together_beta_r}(c) starts to rapidly increase with $e$. 
    For stronger surface attraction, the situation gets more complicated since the adsorption and collapse peaks overlap in Figure \ref{fig:together_beta_r}(a) and (b). 
    For $\epsilon_s=1$ one can at least identify a small adsorption peak for the free chain at $e\approx -1.95$ that clearly differs from the collapse peak,
    but for $\epsilon_s= 2$ and larger the collapse peak disappears and becomes a shoulder at lower $e$ (for $\epsilon_s=5$ at $e\approx -4.75$). 
    This disappearence of the peak is a finite-size effect. For a continuous transition like the collapse transition\cite{deGennes}, one expects a peak of 
    $d\beta(e)/de=-(\partial s(e)/\partial e)^2_{N,V}/c_V(e)$ with a negative maximum\cite{gamma} that we indeed observed for longer chains.
    Again, the positions of the shoulders fall into a regime where $R_{\rm gyr}^2(e)$ strongly increases with $e$.
    Also, canonical data of $d\left<R_{\rm gyr}^2\right>/dT$ confirm that this is indeed the collapse transition 
    because the temperature of the canonical collapse peak is close to the microcanonical temperature that corresponds to this energy.
    In the low-surface-attraction regime, the radius of gyration components parallel and perpendicular to the substrate are similar, such that this maximum can also be seen
    in Figure \ref{fig:rgyrz} and \ref{fig:es0.7}(b) at $T\approx 2$. 
    Regarding the temperature dependence of the collapse transition, it is worth noting that the maximum of the radius of gyration remains at the same temperature until
    collapse and adsorption transitions lines intersect (\ref{fig:diagram}). At higher surface attraction strengths it moves to slightly lower temperatures.
    This reduction of the collapse temperature is rooted in the decreasing ratio of $\left<R_{\rm gyr,\perp}^2\right>/\left<R_{\rm gyr,\parallel}^2\right>$ with increasing $\epsilon_s$: 
    A deformation that effectively decreases the number of possible monomer-monomer contacts and with it the energetical advantage of collapsing. The collapse shifts from three-dimensional
    behavior for weak surface attraction to almost effectively two-dimensional behavior for strong surface attraction.
    This, however, happens for both grafted and nongrafted polymers.

    Now, for energies above the collapse transition the squared radius of gyration in Figure \ref{fig:together_beta_r}(c) for grafted polymers always exceeds that of free polymers.
    This effect gets the strongest at the adsorption transition for strong surface attraction. 
    For $\epsilon_s=3,4,5$ at $e\approx -0.8$, the free chain gets more compact after desorption. Just before it desorbs, it lies quite extended and preferentially flat
    on the substrate in the AE phase. 
    As soon as it leaves the influence of the surface field, this surface-induced deformation vanishes, and the more spherical bulk shape with a lower radius of
    gyration is adapted.
    A grafted polymer cannot leave the surface field and the deformation persists often with a depletion regime\cite{netz} at the substrate.
    It is very likely that this deformation is related to the first-order-like behavior of finite expanded conformations at the adsorption transition. 
    The decreasing $R_{\rm gyr}^2(e)$ values in Figure \ref{fig:together_beta_r}(c) at the adsorption transition fit to the positive values of $d\beta(e)/de$ in Figure \ref{fig:together_beta_r}(b)
    that directly reflect the convex regime in $s(e)$ (where $c_V(e)$ is negative). Neither grafted expanded nor grafted or free collapsed polymers get that significantly deformed during the 
    adsorption process, and in all those cases already the adsorption process of finite polymers is continuous.
    Hence, at phase coexistence the coexisting adsorbed and desorbed phases are separated by a conformational rearrangement.\\

\textbf{Globule Adsorption versus Wetting.}
    For globular chains it even is nontrivial to identify an adsorption transition if the polymer is grafted. 
    A globular chain attached to a substrate always has several surface contacts such that a ``desorbed globule'' ceases to be a well-defined description here.
    One might, however, identify the transition from attached globules that only have a few contacts because the monomer-monomer attraction exceeds the monomer-surface
    attraction, to docked conformations for stronger surface attraction strengths with the wetting transition\cite{wetting}. 
    This roughly coincides with the position of the adsorption transition for the free chain between DG and AG in the phase diagram. 
    For $N=40$, this wetting is not signaled in a pronounced way in our data, but it is visible. 
    Figure \ref{fig:rgyrz} shows the temperature fluctuation of the tensor component of the radius of gyration perpendicular to the substrate $d\left<R_{\rm gyr, \perp}^2\right>/dT$ 
    for (a) the grafted and (b) the nongrafted chain. While in the free case a maximum along the whole line $\epsilon_s\propto T$, with a constant of proportionality 
    close to one, is visible, the activity at the adsorption transition is strongly reduced in the grafted case, and below the collapse transition, no maximum is visible.
    In Figure \ref{fig:es0.7}, some canonical expectation values for a surface attraction in this regime ($\epsilon_s=0.7$) are presented, where we also include data for the 
    number of monomers in contact with the substrate, $n_s$. While for the free polymer the adsorption is signaled in the specific heat $c_V(T)$ (a), the fluctuation of 
    the radius of gyration component perpendicular to the substrate $d\left<R_{\rm gyr,\perp}^2\right>(T)/dT$ (b), and the fluctuation of the number of monomers $d\left<n_s\right>(T)/dT$ (c), 
    for the grafted polymer only the negative peak of $d\left<n_s\right>(T)/dT$ is left. This peak indicates the wetting transition and the missing other signals show the difference to adsorption.
    Wetting is a conformational rearrangement with almost no influence on the position of the polymer.
    It should be mentioned here that the exact definition of a ``surface contact'' has hardly any influence on the peak position of $d\left<n_s\right>(T)/dT$ for a free polymer, but for
    the grafted polymer the influence can be quite strong. As already mentioned, we define a monomer $i$ to be in contact to the substrate if $z_i<1.5$.\\

\textbf{Freezing Transition.}
    Although little affected by the grafting, it is instructive to have a closer look at the freezing transition as well.
    Freezing occurs at a transition energy, below which the number of available states is significantly reduced. Here, for a reduction in energy the system has to pay with a considerable loss of 
    entropy and ``freezes'' into the few remaining conformations. In Figure \ref{fig:together_s_t}(a), one sees that for all $\epsilon_s$ there exists an energetic transition point, where $s(e)=\ln g(e)/N$ strongly
    decreases for a small reduction in $e$. If this energy marks the freezing transition, which is known to be first order for large systems, one would expect the first-order character to be reflected 
    in a backbending in $\beta(e)$ or equivalently a positive maximum in $d\beta(e)/de$.
    Such a backbending as was found for the adsorption of a finite free polymer, should tend to a linear regime in $s(e)$ 
    and a constant regime in $\beta(e)$ for infinite system size.
    For a chain length of $N=40$ no maximum in $d\beta(e)/de$ is visible. Instead, one can see a small shoulder. 
    %
Hence, for small systems, the freezing transition is continuous. 
    At a certain system size, the entropic barrier of cooperative monomer rearrangement within the polymer has increased 
    to such a degree that the system cannot easily go from one phase to the other. This is the same effect that makes it so hard to precisely sample low-energy conformations of large systems.

    Because of the almost identical conformational entropies, the grafting does not affect the freezing transition noticeably. Of course, if the ground state of the free polymer is forbidden 
    by the grafting constraint, it cannot be adapted and the behavior is modified. However, the relative restriction of the entropy below the freezing transition due to grafting is about 
    the same as right above it such that the effect is not relevant at the ensemble level. 

\section{Conclusions}
    In conclusion, we have analyzed and compared the whole phase diagram of a generic off-lattice model for grafted and nongrafted polymer chains for a range of temperatures and surface interaction strengths. 
    The main differences were found at the adsorption transition.
    Here, the restriction of translational entropy above the transition due to grafting is much stronger than below the transition. Additionally, the grafting reduces the 
    necessary rearrangement of segments to form substrate contacts and to adsorb such that grafted adsorption is always continuous while the adsorption of the free chain
    exhibits first-order-like signatures for strong surface attraction and short chains. 
    When for grafted chains, the adsorption temperature is below the coil-globule transition temperature, there are always several surface contacts present and the adsorption changes
    into the wetting transition. For free chains, a continuous adsorption transition exists here. 

    Especially the crossover from a first-order-like to a continuous transition for strong surface attraction
    is in interesting contrast to the freezing transition. This transition is hardly affected by grafting. But despite known to be of first order in the limit of long chains, 
    the microcanonical analysis reveals it to be continuous for short chains. 
    The likely reason is the different nature of the two transitions. While the freezing transition -- like the coil-globule transition -- effectively requires a rearrangement 
    of the individual monomers relative to each other, what requires more energy for longer chains and offers more possibilities for barriers between different conformations,
    the adsorption transition depends on the center-of-mass of the whole polymer and energy gained by all surface contacts. 

\section{Acknowledgment}
    This work is partially supported by the DFG (German Science 
    Foundation) under Grant Nos.\ JA \mbox{483/24-1/2/3}, the Leipzig 
    Graduate School of Excellence GSC 185 ``BuildMoNa'', the SFB/TRR 102, 
    the Deutsch-Franz\"osische Hochschule (DFH-UFA) under Grant No. CDFA-02-07,
    and by the German-Israel Program ``Umbrella''
    under Grant Nos.\ SIM6 and HPC\_2. 
    Support by supercomputer time grants (Grant Nos.~hlz11, hlz17, JIFF39 and JIFF43) of the 
    Forschungszentrum J{\"u}lich is gratefully acknowledged.

\end{document}